\documentclass[aps,prl,twocolumn,showpacs,superscriptaddress,amsmath,amssymb]{revtex4-1}
\usepackage{graphicx}
\usepackage{color,soul}
\usepackage{xcolor}

\begin{document}

   \title{Structural criterion for the onset of rigidity in a colloidal gel}
     
\author{C. S. Dias} \email{csdias@fc.ul.pt}
\affiliation{Centro de F\'{i}sica Te\'{o}rica e Computacional, Faculdade de Ci\^{e}ncias, Universidade de Lisboa, 1749-016
Lisboa, Portugal} 
\affiliation{Departamento de F\'{\i}sica, Faculdade de Ci\^{e}ncias,
Universidade de Lisboa, 1749-016 Lisboa, Portugal} 

\author{J. C. Neves}\email{fc51402@alunos.fc.ul.pt}
\affiliation{Centro de F\'{i}sica Te\'{o}rica e Computacional, Faculdade de Ci\^{e}ncias, Universidade de Lisboa, 1749-016
Lisboa, Portugal} 
\affiliation{Departamento de F\'{\i}sica, Faculdade de Ci\^{e}ncias,
Universidade de Lisboa, 1749-016 Lisboa, Portugal} 

\author{M. M. Telo da Gama} \email{mmgama@fc.ul.pt}
\affiliation{Centro de F\'{i}sica Te\'{o}rica e Computacional, Faculdade de Ci\^{e}ncias, Universidade de Lisboa, 1749-016
Lisboa, Portugal}
\affiliation{Departamento de F\'{\i}sica, Faculdade de Ci\^{e}ncias,
Universidade de Lisboa, 1749-016 Lisboa, Portugal}

\author{E. Del Gado} \email{emanuela.del.gado@georgetown.edu}
\affiliation{Department of Physics, Institute for Soft Matter Synthesis and Metrology, Georgetown University, Washington, D.C. 20057, USA}

\author{N. A. M. Ara\'ujo} \email{nmaraujo@fc.ul.pt}
\affiliation{Centro de F\'{i}sica Te\'{o}rica e Computacional, Faculdade de Ci\^{e}ncias, Universidade de Lisboa, 1749-016
Lisboa, Portugal}
\affiliation{Departamento de F\'{\i}sica, Faculdade de Ci\^{e}ncias,
Universidade de Lisboa, 1749-016 Lisboa, Portugal} 

  \begin{abstract}
Identifying the necessary conditions for the onset of rigidity in a gel
remains a challenge. It has been suggested that local particle coordination
could be used to establish such conditions, but rigid gels occur for various
coordination numbers. Combining simulations, oscillatory rheology, and a
percolation analysis, for particles where the valence can be controlled, we
find that the onset of rigidity coincides with the percolation of particles
with three or more bonds, which arises after the connectivity percolation. We
show that the rigidity results from an interplay of bonding and non-bonding
interactions, providing insight into low-valence colloidal gel rigidity.
  \end{abstract}

  \maketitle

Colloidal gels are non-equilibrium structures sustained by a percolating
network of long-living bonds between the particles in
solution~\cite{Lu2008,Pusey1986,Bonn1999,Klix2010,Puertas2002}. They exhibit
unique
structural~\cite{Ferreiro-Cordova2020,Griffiths2017,Colombo2013,Machlus2021},
dynamical~\cite{Cho2020,Tavares2018,Bonacci2020}, and
mechanical~\cite{Zhang2019,Colombo2014a,Tsurusawa2020} properties, which have
motivated an intensive search for strategies to design them. This effort is
relevant to a range of industrial applications, from food to cosmetics, and
from energy materials to biotechnology and construction
\cite{Mitura2020,Ioannidou2016,Banerjee2012,Coropceanu22,Dias2020a}.
Recently, one focus has been on the design of low density gels and in the
control of their local
structures~\cite{Ruzicka2011,Bianchi2006,Smallenburg2013,Pawar2010,Lee2011,He2012,Wang2015,Shah2013,Bianchi2015,Romano2011,Markova2014,Ortiz2014,Wolters2015}.
This may be achieved using, for example, vitrimetric materials or patchy
particles, which limit the valence of the particles and the directionality of
the bonds~\cite{Smallenburg2013a,Lei2020,Capelot2012,Liu2020a}, helping to
design the gel structures.

The key property of colloidal gels is their capacity to bear load without
deforming permanently. However, the necessary condition for a gel to be rigid
is still under debate~\cite{Fenton2023}. This question is even more critical
in the design of low-density gels, where the reduced number of bonds per
particle may compromise the mechanical stability. During gelation, bonds are
established between pairs of particles until a percolating network of bonds
emerges at a critical fraction of bonds per
particle~\cite{Lu2008b,Pusey1986,Bonn1999,Klix2010,Puertas2002}. At the onset
of this connectivity percolation, the gels are floppy and
fragile~\cite{Kohl2016,Valadez-Perez2013,Richards2017}. As more bonds are
formed, a rigid elastic network emerges at a rigidity percolation
transition~\cite{SampaioFilho2018,Zhang2019}. The necessary conditions for the
rigidity percolation are unclear, since it is difficult to identify which
subset of the visible structure is responsible for the stress
transmission~\cite{Zhang2019}. In 2D the onset of rigidity is identified
directly from the bond network with the pebble game devised for rigidity
percolation~\cite{Jacobs1995,Ellenbroek2015,Berthier2019}, but in 3D the
criteria remain elusive. Several local and global geometrical criteria have
been proposed, such as a minimum average coordination
number~\cite{Valadez-Perez2013} or the existence of a directed percolating
path~\cite{Kohl2016}, but a recently correlated rigidity percolation framework
highlights the non-local nature of the emergence of rigidity, casting doubts
on the previously proposed criteria~\cite{Zhang2019}.

Experiments combining confocal microscopy with microrheology in colloidal gels
resulting from depletion forces suggest that the onset of rigidity coincides
with the percolation of particles with at least six
neighbors~\cite{Tsurusawa2019}, in line with a local Maxwell rigidity
criterion for isostaticity~\cite{Blumenfeld2005}. However recent experiments
and simulations have clarified that this local criterion does not provide the
necessary conditions for gel rigidity~\cite{Fenton2023}. While attractive
depletion forces are isotropic and pairs of bonded particles are free to
rotate relative to each other~\cite{Blumenfeld2005}, rotational degrees of
freedom can be hindered by molecules grafted on the particle
surfaces~\cite{Feng2013,Sacanna2013,Wang2012} and, quite generally, by more
complex surface interactions~\cite{Bantawa2021,Bonacci2020,VanderMeer2022}.

Experimentally, the count of neighbors is based on the number of particles
within a certain \textit{bonding} distance, which becomes problematic when
different types of interactions are present. Here, we perform numerical
simulations for gels consisting of particles with fixed valence, allowing us
to distinguish bonds from contacts. We show that this difference is not a mere
detail. For gels of limited valence, rigidity does emerge from the interplay
of bonds and repulsive (contact) interactions.  We find that the onset of
rigidity coincides with the emergence of a spanning aggregate of particles
with at least three bonds for any value of the valence, including the
isotropic continuum limit. This poses challenges for the development of
algorithms like the \textit{pebble game} which aim at identifying the onset of
rigidity by decomposing the network into local features~\cite{Jacobs1995}. 

Colloidal particles with limited valence spontaneously form gels of very-low
density~\cite{Russo2009,Araujo2017,DelasHeras2012,Russo2011a,Russo2022}, with
the focus of most theoretical and numerical works being the connectivity
percolation~\cite{Howard2019,Zaccarelli2005,Zaccarelli2006,Ruzicka2011,Bianchi2011}.
However, for most practical applications, it is critical that the gel is
rigid. From the findings reported here, it is clear that the relevant
structural criterion is not the onset of connectivity percolation but rather
the percolation of particles with at least three bonds. This is a significant
difference since the latter typically occurs at much later times, due to the
slow aging of the gel~\cite{Dias2018b,Colombo2014,Zhang2019,Fenton2023}.
Specific local geometric motifs, which may depend on the particle
interactions, contribute to attaining rigidity, however it is their spatial
organization and the way they are embedded in the structure that drives the
onset of rigidity.  Our results are consistent with, and provide further
insight into, recent studies of nanoparticle gel
rigidity~\cite{Fenton2023,Bantawa2023}.

\begin{figure}[t]
   \begin{center}
\includegraphics[width=1\columnwidth]{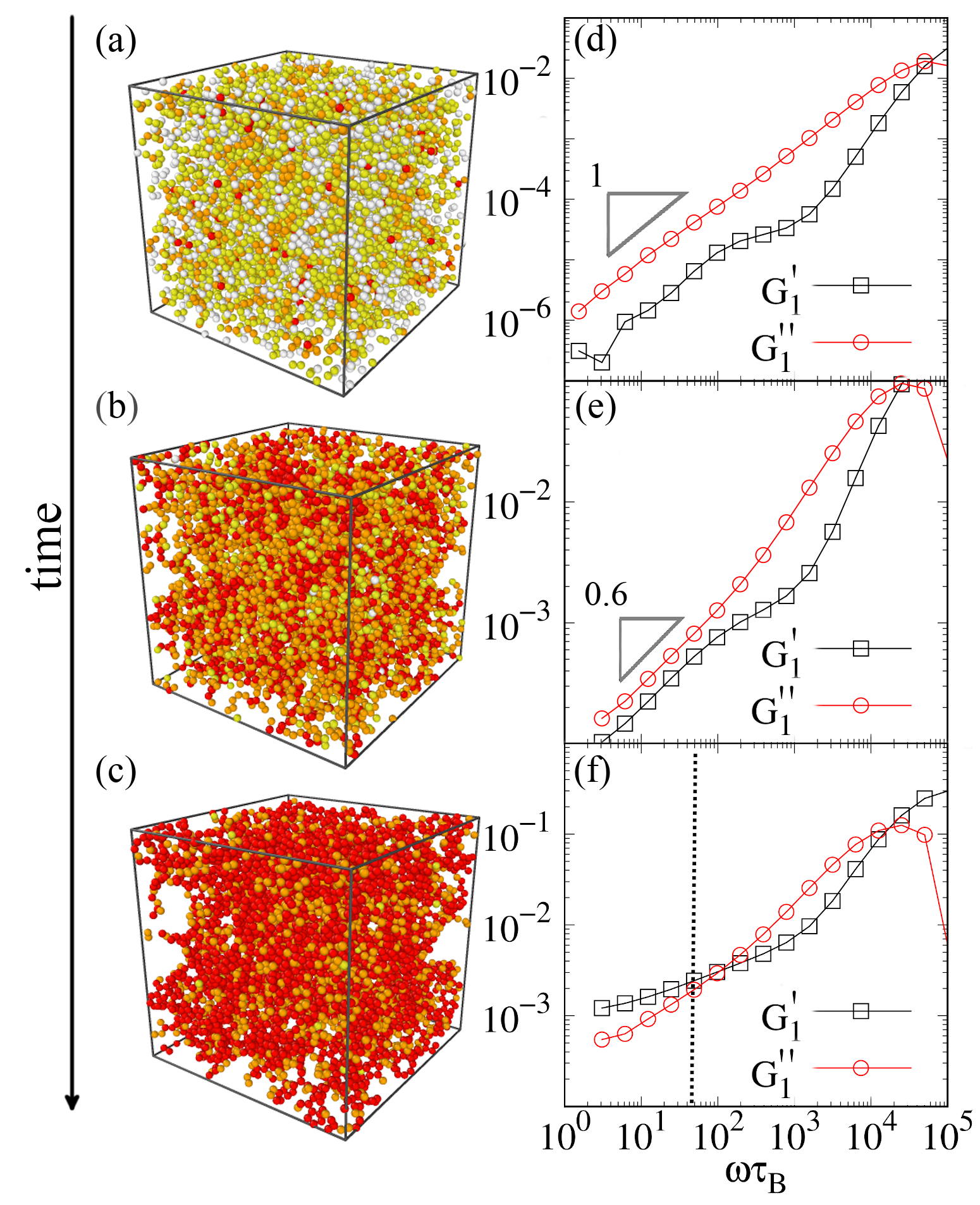} \\
\caption{\textbf{Time-dependence of the mechanical properties of a gel of
spherical particles with valence three.} (a)-(c) three snapshots
at different times. The color of each particle indicates its number of bonds:
white (0), yellow (1), orange (2), and red (3). (d)-(f) the
storage ($G^{\prime}_1$) and loss ($G^{\prime \prime}_1$) moduli at the same
times. (a) and (d) correspond to a viscoelastic fluid, (b) and (e) to a soft
glassy material, and (c) and (f) to an elastic gel.  Mechanical properties
were measured in one sample of a cubic box of linear size $L=32$, in units of
the particle diameter, and number density $\rho=0.2$.} \label{fig.mech_time}
 \end{center}
\end{figure}

\begin{figure}[t]
   \begin{center}
\includegraphics[width=0.9\columnwidth]{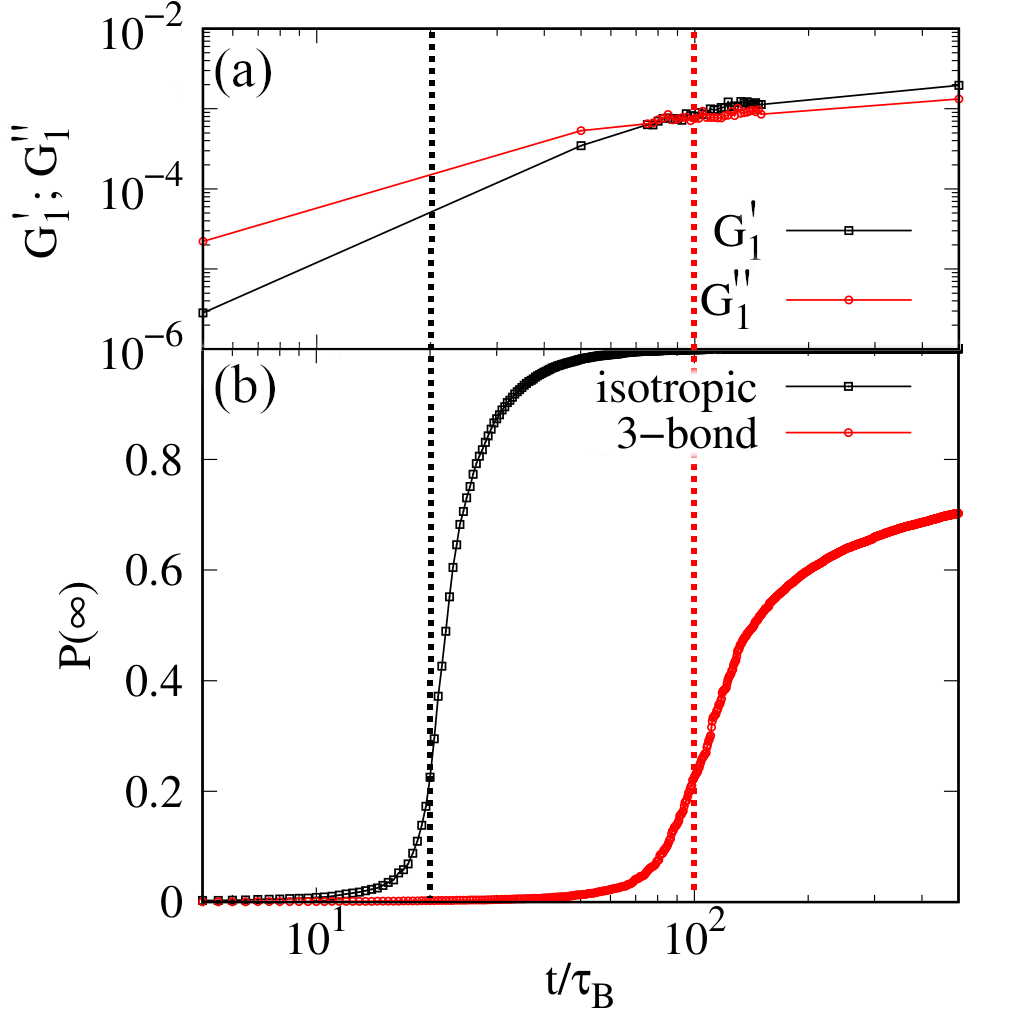} \\
\caption{\textbf{Mechanical and percolation properties of a gel of spherical
particles with valence three.} (a) Storage $G^{\prime}_1$ and loss
$G^{\prime\prime}_1$ moduli as a function of time, measured at a frequency
corresponding to the vertical line in Fig.\ref{fig.mech_time}(f). (b) Fraction
of particles, $P(\infty)$, in the largest connected component (black) and
largest cluster of particles with three bonds (red) as a function of
time. Vertical black (left) and red (right) dashed lines indicate the
threshold for the connectivity and three-bond percolation transitions,
respectively. All simulations were performed on a cubic box of linear size
$L=32$, in units of the particle diameter, number density $\rho=0.2$, and the
results are averaged over 10 samples for the percolation analysis and
correspond to only one sample for the mechanical
properties.~\label{fig.oscill_time_mech}}
   \end{center}
\end{figure}

\textit{Model.} We consider a monodisperse suspension of spherical colloidal
particles of radius $R$. The valence of each particle is limited by adding
pointlike attractive sites on the surface of the particles and considering
that bonds are only established through the attractive sites. Thus, the number
of attractive sites sets the valence. The pairwise interaction is then a
superposition of an isotropic repulsion and a site-site attraction. As in
previous works~\cite{Dias2016,Dias2018b}, the repulsive interaction is given
by,
\begin{equation}
U_{\text{colloid/colloid}}(r)=A e^{-\left(r-2R\right)/k}, \label{eq.yukawa}
\end{equation}
where $r$ is the distance between the center of the particles, $A=1$ provides
the energy scale, and $k=4$ is the screening length in reduced units. The
bonding interaction is given by an inverted Gaussian potential, 
\begin{equation}
U_{\text{site/site}}(r_s)=-\epsilon e^{-(r_s/\sigma)^2}, \label{eq.gaussian} 
\end{equation}
where $r_s$ is the distance between the attractive sites, $\epsilon=20$ is the
interaction strength, and $\sigma=0.1R$ the width of the Gaussian in reduced
units. The trajectories were obtained numerically by Langevin dynamics, using
the LAMMPS~\cite{Plimpton1995}, for diffusion coefficients for the translation
$D_\mathrm{t}$ and rotational $D_\mathrm{r}$ motion that are related by the
Debye-Einstein relation
$D_\mathrm{r}=\frac{3}{4R^2}D_\mathrm{t}$~\cite{Mazza2007}.

We first consider particles with three attractive sites on a cubic box of
linear size $L=32$ in units of the particle diameter and periodic boundary
conditions in all directions. To form an elastic gel, the number density is
set to $\rho=0.2$ and the temperature is $k_\mathrm{B}T/\epsilon=0.025$.  The
dynamics evolves over a total time of $10^4\tau_\mathrm{B}$, where
$\tau_\mathrm{B}=4R^2/D_\mathrm{t}$ is the Brownian time.

\textit{Rheological measurements.} We minimize the energy by running
zero-temperature overdamped simulations until the kinetic energy is, at least
ten orders of magnitude lower than its initial value~\cite{Colombo2014}. Then,
we apply Lees-Edwards boundary conditions and an affine shear deformation,
\begin{gather}
 \mathbf{r'_\mathit{i}}=
  \begin{bmatrix}
   1 & \gamma(t) & 0 \\
   0 & 1 & 0 \\
   0 & 0 & 1 
   \end{bmatrix}
   \mathbf{r}_i, 
\end{gather}
where $\textbf{r}_i$ and $\textbf{r}'_i$ are the initial and final positions
of particle $i$, and $\gamma(t)$ is the shear strain~\cite{Colombo2014}.
$\gamma(t)$ oscillates periodically in time, $\gamma (t)=\gamma_0 \sin(\omega
t)$, with frequency $\omega$ and amplitude $\gamma_0$. We computed the load
curve to identify the linear regime. To focus on that regime, we set
$\gamma_0=0.07$. The first-order storage $G^{\prime}_1$ and loss
$G^{\prime\prime}_1$ moduli are obtained from the time evolution of the shear
stress $\sigma_{xy}(t)$ as in Ref.~\cite{Colombo2014}.

\begin{figure}[t]
   \begin{center}
\includegraphics[width=0.9\columnwidth]{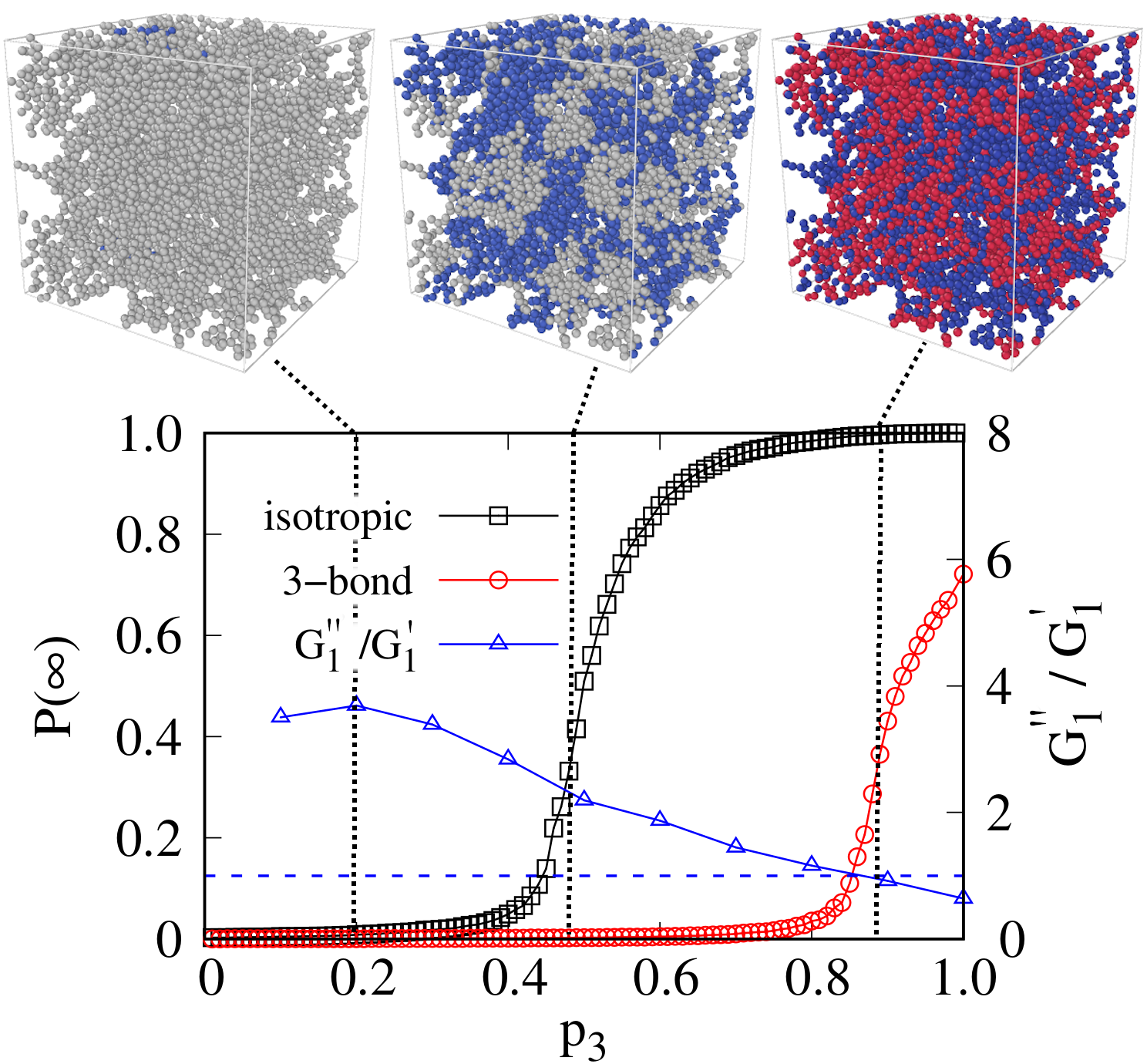} \\
\caption{\textbf{Percolation analysis of the mechanical properties of a gel of
spherical particles with valence three, where sites are removed at random.}
Fraction of particles, $P(\infty)$, in the largest connected component (black
squares) and largest cluster of particles with three bonds (red squares) as a
function of the fraction of the remaining particles with three attractive
sites $p_3$. Loss factor (blue triangles)
$\tan(\delta)=G^{\prime\prime}_1/G^{\prime}_1$ as a function of $p_3$. Top
panel: Snapshots at three different stages (vertical dashed lines in the
plot). Red particles are in the largest cluster of particles with three bonds
(fully bonded), blue particles are in the largest connected component, and all
the other particles are gray. All simulations were performed on a cubic box of
linear size $L=32$, in units of the particle diameter, for a number density
$\rho=0.2$. The results are averaged over 10 samples for the percolation
analysis and correspond to only one sample for the mechanical properties.
\label{fig.3patch_remove}}
   \end{center}
\end{figure}

\begin{figure}[t]
   \begin{center}
\includegraphics[width=0.9\columnwidth]{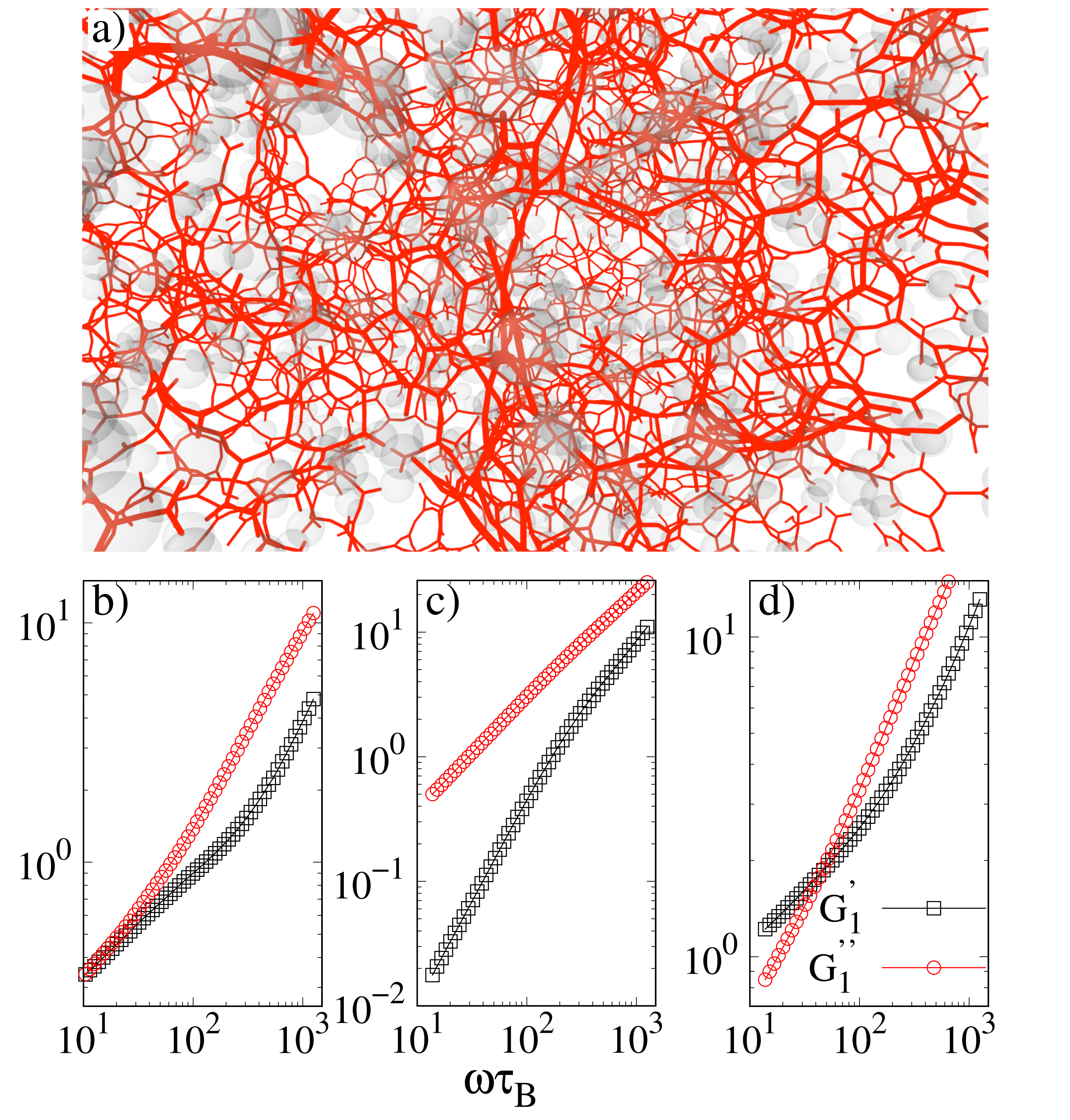} \\
\caption{\textbf{Structure and mechanical properties of the cluster of
three-bonded particles.} (a) Snapshot of the largest connected component at
the onset of rigidity, where the cluster of particles with three bonds is
represented by the network of bonds (red) and all the other particles are in
gray. (b), (c), and (d) are the storage $G^\prime_1$ and loss
$G^{\prime\prime}_1$ moduli as a function of the frequency $\omega$. (b)
depicts the cluster of particles with three bonds (red network), (c) the
largest connected component when we switch off the repulsive interactions
during the rheological measurements, and (d) the largest connected component
when we switch off the repulsive interaction only between the particles with less
than three bonds during the rheological measurements.  All simulations were
performed on a cubic box of linear size $L=32$, in units of particle diameter,
and number density $\rho=0.2$, for one sample.}
\label{fig.3bondcluster}
   \end{center}
\end{figure}

Figures~\ref{fig.mech_time}(a)-(c) are snapshots of the colloidal system at
different times, where the color of a particle corresponds to the number of
connected neighbors. Figures~\ref{fig.mech_time}(d)-(f) show the $\omega$
dependence of $G^{\prime}_1$ and $G^{\prime\prime}_1$ for the three different
regimes of gelation.  Initially, in (a) and (d), most particles have
established either one or no bonds. $G^\prime_1$ is lower than
$G^{\prime\prime}_1$ and it scales linearly with $\omega$, as in a Maxwell-like
fluid~\cite{Mason1995}. As time evolves, more bonds are established and
aggregates are formed with a characteristic size of several particles,
affecting the mechanical response.  For (b) and (e), $G^{\prime}_1$ is still
lower than $G^{\prime\prime}_1$ but both scale as $\omega^{\alpha}$, with
$\alpha<1$; We estimate $\alpha=0.60\pm0.05$~\cite{Bantawa2023}. Finally, in
(c) and (f), a percolating network of particles with three bonds is already
formed and the gel is rigid, since $G^{\prime}_1>G^{\prime\prime}_1$ at low
enough frequencies.

The time dependence of $G^\prime_1$ and $G^{\prime\prime}_1$ for a frequency
$\approx 50~\tau_\mathrm{B}^{-1}$ is shown in
Fig.~\ref{fig.oscill_time_mech}(a) (vertical line in
Fig.\ref{fig.mech_time}(f)).  Both moduli increase with time and the
transition to an elastic gel ($G^\prime_1>G^{\prime\prime}_1$) occurs around
$100\tau_\mathrm{B}$ (vertical red dashed line). The fraction of particles in
the largest connected component is plotted in
Fig.~\ref{fig.oscill_time_mech}(b) (black curve).  The onset of rigidity
occurs much later than that for the onset of connectivity percolation
(vertical black dashed line, from Ref.~\cite{Dias2018}).  However, if we
compute the size of the largest cluster of fully-bonded particles (red curve),
the data indicate that the onset of percolation of this subset 
coincides with the onset of rigidity.

To further test if the onset of rigidity coincides with the percolation
threshold of particles with three bonds, we performed the
following percolation analysis. We started from gel configurations at the
final time of the simulations, when the gel is
predominantly elastic and the particles with three bonds percolate. Then, at
each iteration, one particle is selected and one of its attractive sites is
removed. Both the particle and the attractive site are selected at random. We
evaluate the mechanical and percolation properties as a function of $p_3$,
defined as the fraction of particles that still have three attractive sites.
Figure~\ref{fig.3patch_remove} depicts the dependence on $p_3$ of the size of
the largest connected component, the largest cluster of particles with three
bonds and the ratio $G^{\prime}_1/G^{\prime\prime}_1$.  The data reveals that
the onset of rigidity and the percolation of particles with three bonds
coincide.  Note that particles and attractive sites are selected uniformly at
random, discarding possible effects from spatial correlations developed during
the dynamics.

A snapshot of the largest connected component at the onset of rigidity is
shown in Fig.~\ref{fig.3bondcluster}(a), where the set of fully-bonded
particles is represented by the network of bonds (red). The cluster of
three-bonded particles forms a spanning network with cycles of at least five
particles. Surprisingly, this structure is not rigid in itself, as shown in
Fig.~\ref{fig.3bondcluster}(b), where we plot $G^\prime_1$ and
$G^{\prime\prime}_1$ for that structure, when, during the rheological
measurements, we neglect the interaction with all the other particles. Even
when we consider the largest connected component (all particles), switching
off the repulsive interactions between the particles, while keeping the
bonding interaction, the structure is not rigid (see
Fig.~\ref{fig.3bondcluster}(c)).  It is only when we include the repulsive
interactions, at least between the particles with three bonds, that the
structure becomes rigid (see Fig.~\ref{fig.3bondcluster}(d)). Thus, although
the onset of rigidity coincides with the percolation of particles with three
bonds, it emerges from a combination of those bonds and repulsive (steric)
interactions. These findings suggest that the rest of the structure in which
the bond network is embedded contributes to rigidity.

\begin{figure}[t]
   \begin{center}
\includegraphics[width=1\columnwidth]{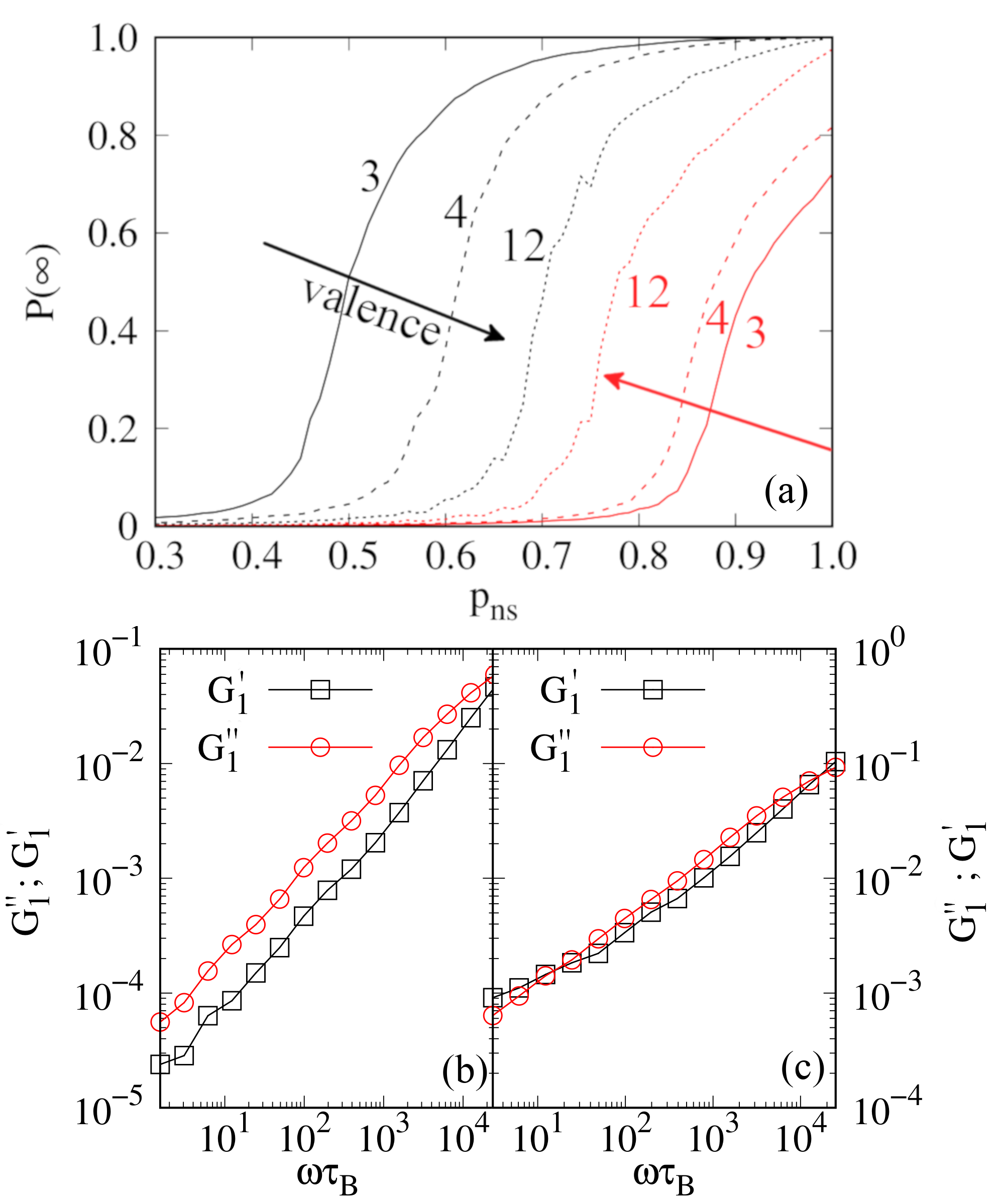} \\
\caption{\textbf{Dependence on the valence.} (a) Fraction of particles,
$P(\infty)$, in the largest connected component (black) and largest cluster of
particles with at least three bonds (red) as a function of the fraction of
particles with $ns$ attractive sites $p_{ns}$, where $ns$ is the valence.
Different curves are for different values of $ns$, namely, three (solid), four
(dashed), and twelve (dotted).  (b) and (c) Storage $G^\prime_1$ and loss
$G^{\prime\prime}_1$ moduli as a function of the frequency $\omega$ for a
fraction of particles with twelve attractive sites (b) $p_{12}=0.6$ and (c)
$p_{12}=0.8$, which are below and above the onset of percolation for particles
with at least three bonds (see red dotted curve in (a)). All simulations were
performed on a cubic box of linear size $L=32$, in units of particle diameter,
number density $\rho=0.2$, and the results are averaged over 10 samples in
(a) and are for one sample in (b) and (c).} \label{fig.transition_valence}
   \end{center}
\end{figure}

\textit{Effect of the valence.} So far, only particles of valence three have
been considered. To evaluate the dependence on the valence, we performed the
same percolation analysis for gels of particles with valence $ns$ up to
twelve, which is the number of neighbors in compact 3D packings. We
select particles at random and remove all of their attractive sites but two.
Thus, at each iteration the particles have either $ns$ or two attractive
sites. In Fig.~\ref{fig.transition_valence}, the size of the largest connected
component (black) and the largest cluster of particles with at least three
bonds (red) are plotted as a function of the fraction $p_{ns}$ of particles
with $ns$ attractive sites. The onset of the two percolation transitions
depends strongly on the valence $ns$. While the one for connectivity
percolation increases with $ns$, that for particles with at least three bonds
shifts to lower values of $p_{ns}$. Thus, for particles with higher valence
the two transitions are very difficult to distinguish experimentally.

Figure~\ref{fig.transition_valence} also shows $G^\prime_1$ and
$G^{\prime\prime}_1$ as a function of $\omega$ for a gel where $ns=12$.  We
consider two different configurations (b) $p_{12}=0.6$ and (c) $p_{12}=0.8$,
which are below and above the onset of percolation for particles with at least
three bonds. As in the case of $ns=3$ (Fig.~\ref{fig.3patch_remove}), the
$G^{\prime}_1$ is only larger than $G^{\prime\prime}_1$ above the percolation
transition for particles with at least three bonds. Below this transition,
$G^{\prime\prime}_1$ is larger, over the entire range of frequencies, and thus
the gel is not elastic but rather behaves as a viscous fluid.

\textit{Conclusion.} We found that it is the combination of a bond network and
repulsive contacts that leads to the rigidity of the gel. The development of
rigid structures depends strongly on the gelation dynamics, not just on the
number of bonds of a particle, as proposed by correlated rigidity
percolation~\cite{Zhang2019}. For all the gels considered here, where the
particle valence ranged from three to twelve, we found that the onset of gel
rigidity corresponds to the percolation of particles with at least three
bonds, suggesting that for the type of interactions considered, this
identifies the necessary condition for rigidity percolation. For nanoparticles
with ad-hoc limited valence, our results provide the insight that a sufficient
fraction of particles with valence three is required to design rigid
structures. More generally, when the valence can not be controlled precisely,
our findings shed light into how the mechanics of colloidal gels results from
the interplay of different substructures intertwined with each other, not
easily recognizable from local geometrical motifs.

\textit{Acknowledgments.}
We acknowledge financial support from the Portuguese
Foundation for Science and Technology (FCT) under Contracts no.
PTDC/FIS-MAC/28146/2017 (LISBOA-01-0145-FEDER-028146), PTDC/FIS-MAC/5689/2020,
EXPL/FIS-MAC/0406/2021, CEECIND/00586/2017, UIDB/00618/2020, and
UIDP/00618/2020. EDG acknowledge financial support from NSF DMR-2026842 and
ACS Petroleum Fund. The authors also thank Minaspi Bantawa for support with
the rheological tests.

\bibliography{library}

\end{document}